\documentclass[preprint,showpacs,preprintnumbers,amsmath,amssymb]{revtex4}
\usepackage{epsfig}
\usepackage{graphicx}
\usepackage{ulem}
\usepackage{color}
\definecolor{My_red}{cmyk}{0.00,1.00,1.00,0.20}

\def\be{\begin{equation}}
\def\ee{\end{equation}}
\def\bea{\begin{eqnarray}}
\def\eea{\end{eqnarray}}
\def\nn{\nonumber}
\def\spur{\not \!}
\begin{document}
\title{Complete Analysis on the Short Distance Contribution of
$B_s\to \ell^+\ell^-\gamma$ in Standard Model}
\author{Wenyu Wang, Zhao-Hua Xiong and Si-Hong Zhou}
\affiliation{Institute of Theoretical Physics, College of Applied Science,
     Beijing University of Technology, Beijing 100124, China}
\date{\today}
\begin{abstract}
Using the $B_s$ meson wave function extracted from non-leptonic $B_s$ decays,
we evaluate the short distance contribution of rare decays 
$B_s\to \ell^+\ell^-~\gamma~~(\ell=e,\mu)$
in the standard model, including all the possible diagrams.
We focus  on the contribution from four-quark operators which are not
 taken into account properly in previous researches.
We found that the contribution is  large, leading to  the branching ratio of
 $B_s\to \ell^+\ell^-\gamma$ being  nearly enhanced by a factor 3 and up to $1.7\times 10^{-8}$. 
 The predictions for such processes can be tested in the LHC-b and B factories in near future.
\end{abstract}
\maketitle

\section{Introduction}
The standard model (SM) of electroweak interaction has been
remarkably successful in describing physics below the Fermi
scale and is in good agreement with the most experiment data.
One of the most promising processes for probing the quark-flavor sector
of the SM is the rare decays. These decays, induced
by the flavor changing neutral currents (FCNC) which occur in
the SM only at loop level, play an important role in the phenomenology
of particle physics and in searching for the physics beyond the SM
\cite{Aliev97,Xiong01}. The observation of the penguin-induced decay
$B\to X_s\gamma$,~$B\to X_s\ell^+\ell^-(\ell=e,\mu)$ are in
good agreement with the SM prediction, and the first evidence for the decay $B_s \to \mu^+ \mu^-$
was  confirmed at the end of 2012 \cite{Aaij:2012nna}, putting strong constraints on its various
extensions.  Nevertheless, these processes are also important in determining
the  parameters of the SM and some hadronic parameters in QCD,
such as the CKM matrix elements, the meson decay constant $f_{B_s}$,
providing information on heavy meson wave functions.

Thanks to the Large Hadron Collider (LHC) at CERN
we have entered a new era of particle physics. In experimental side, in the current early
phase of the LHC era, the exclusive modes such as $B_s\to\ell^+\ell^-\gamma$ $(\ell=e,~\mu)$ 
 are among the most promising decays due to their relative 
cleanliness and sensitivity to models beyond the SM \cite{Aliev97,Xiong01}. In theoretically side,
since no helicity suppression exists and  large  branching ratios
as $B_s\to \mu^+\mu^-$ are expected.   There are mainly two kinds of 
contributions of $B_s\to\ell^+\ell^-\gamma$ in the SM:
the short distance contribution which can be evaluated reliably by perturbation theory 
\cite{buras}  while the long distance QCD effects describing
the neutral vector-meson resonances $\phi$ and $J/\Psi$ family \cite{Melikhov04,Kruger03,Nikitin11}. 
As for the short distance contribution, it is thought in previous works that 
a necessary work is only attaching real photon to any charged internal and external lines in 
the Feynman diagrams of $b\to s\ell^+\ell^-$ with statement that contributions  from the attachment of 
photon to any charged internal propagator are regraded as to be strongly suppressed and can be neglected
safely~ \cite{Aliev97,Xiong01,LU06,Eilam97}, i.e., one can easily obtain the amplitude of  $B_s\to\ell^+\ell^-\gamma$  by using  the effective weak Hamiltonian of $b\to s\ell^+\ell^-$ and the matrix elements $\langle\gamma|\bar{s}O_ib|B_s\rangle$  $O_i=\gamma_\mu P_L, \sigma_{\mu\nu}q^\nu P_R$ directly.
 Therefore  contributions from the attachment of
real photon with magnetic-penguin vertex to any charged  external lines are of course omitted \cite{Aliev97,Xiong01} or stated  to be negligibly small~\cite{LU06}.
Another contribution from loop insertion of the lower order four-quark operators
are also always neglected.  We note that the complete contribution seems to have been
done in \cite{Melikhov04}, however
it mainly concentrated on the long distance effects of the meson resonances,
whereas the short distance contribution was indeed incompletely analyzed.
A complete examination included all contribution to the processes in the SM is needed.

As being well known, only short-distance contribution can be reliably predicted,
and it is more important than the long-distance contribution from the resonances which
is actually excluded partly by setting cuts in experimental measurements.
Recently we showed that the contributions from the attachment of
real photon with magnetic-penguin vertex to any charged external lines  can enhance the branching ratios
of $B_s\to\ell^+\ell^-\gamma$  by a factor about 2  \cite{Wang:2012na}.

In this letter,  we will extend our previous studies
and  use the $B_s$ meson wave function extracted from non-leptonic $B_s$ decays  \cite{bs}
 to revaluate the short distance contribution from the all categories of diagrams of
$B_s\to \ell^+\ell^-\gamma$ decays. Special attention will be payed on  the contribution
from the four-quark operators, and a comparative study with previous work will be discussed.
The paper are organized in the following, in sec. \ref{sec:hami}, we
analyse the full short distance contribution and present detail
calculation of exclusive decays $B_s\to\ell^+\ell^-\gamma$.
The numerical results and the comparative study are given in sec. \ref{sec:num},
and the conclusions are given in sec. \ref{sec:con}.

%%%%%%%%%%%%%%%%%%%%%%%%%%%%%%%%%%%%%%%%%%%%%%%%%%%%%%%%%%%%%%%%%%%%%%%%%%%%%%%%%%%
\section{Complete analysis on short distance contributions}\label{sec:hami}
In order to simplify the  decay amplitude for $B_s\to \ell^+\ell^-\gamma$,
we have to utilize the $B_s$ meson wave function, which is not known from 
the first principal and model depended. Fortunately, many studies on non-leptonic $B$
\cite{bdecay,cdepjc24121} and $B_s$ decays \cite{bs} have constrained
the wave function strictly.  It was found that the wave function has form
 \begin{equation}
\Phi_{B_s}= (\not \! p_{B_s} +m_{B_s}) \gamma_5 ~\phi_{B_s}
({x}), \label{bmeson}
\end{equation}
where the distribution amplitude $\phi_{B_s}(x)$ can be expressed as~\cite{form}:
\begin{equation}
\phi_{B_s}(x) = N_{B_s} x^2(1-x)^2 \exp \left( -\frac{m_{B}^2\ x^2}{2
\omega_{b_s}^2}  \right)
\label{phib}
\end{equation}
with $x$ being  the momentum fractions shared
by $s$ quark in $B_s$ meson.
 The  normalization constant $N_{B_s}$ can be determined by comparing
\begin{eqnarray}
\langle 0\left|\bar s \gamma^{\mu}\gamma_{5}b\right|B_s\rangle=iN_c\int_{0}^{1}\phi_{B_s}(x)dx
{\rm Tr}\left[\gamma^{\mu}\gamma_{5}(\spur p_{B_s}+m_{B_s})\gamma_{5}\right]dx
=-4N_c ip_{B_s}^{\mu}\int_{0}^{1}\phi_{B_s}(x)dx
\end{eqnarray}
with $N_c$ being the number of quarks and
\begin{eqnarray}
\langle0\left|\bar s\gamma^{\mu}\gamma_{5}b\right|B_s\rangle=-if_{B_s}p_{B_s}^{\mu},
\end{eqnarray}
the  $B$ meson decay constant $f_{B_s}$ is thus determined by
the condition
\begin{eqnarray}
\int_{0}^{1}\phi_{B_s}(x)dx=\frac{1}{4N_c}f_{B_s}.
\end{eqnarray}

\begin{figure}[htbp]
%\vspace{-0.5cm}
\begin{center}
\scalebox{0.8}{\epsfig{file=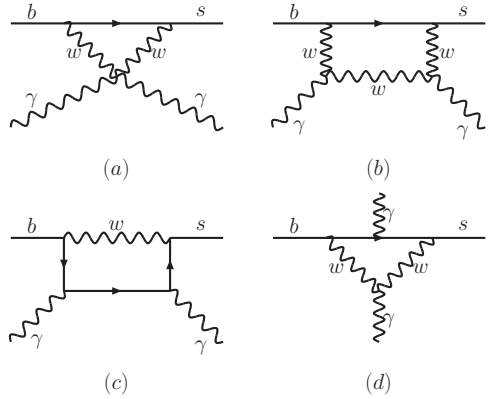}}
\caption{Feynman diagrams without effective vertex of $b\to s\gamma$
 contribution to $B_s\to\gamma\gamma$.}
\label{fig1}
\end{center}
\end{figure}
\begin{figure}[htbp]
%\vspace{-0.5cm}
\begin{center}
\scalebox{0.8}{\epsfig{file=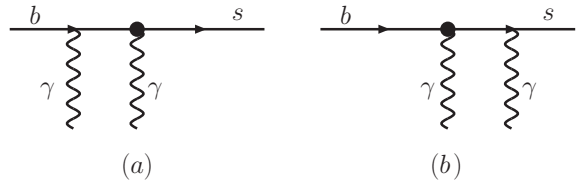}}
\caption{Feynman diagrams with effective vertex of
$b\to s\gamma$ contribution to $B_s\to\gamma\gamma$ .
The black dot stands for the magnetic-penguin operator $O_7 $}
\label{fig2}
\end{center}
\end{figure}
Let us start with the quark level processes $B_s\to \ell^+\ell^-\gamma$
which are subject to the QCD corrected effective weak Hamiltonian.
The general effective Hamiltonian that
describes $b\to s$ transition is given by
\begin{eqnarray}
{\cal H}_{eff}=-\frac{4G_F}{\sqrt{2}}V_{tb}V^*_{ts}\sum\limits_{j=1}^{10}C_j(\mu)O_j(\mu),
\label{hami}
\end{eqnarray}
where $O_j$ ($j=1,\dots,6$) stands for the  four-quark operators, and the forms and
the corresponding  Wilson coefficients $C_i$ can be found in Ref.~\cite{Misiak93}.

Generally,  to describe all the short distance of the process $B_s\to\ell^+\ell^-\gamma$,
new effective operators  for $b\to s\gamma\gamma$ which are not included in (\ref{hami}) should be introduced. Corresponding feynman  diagrams without and with effective vertex $b\to s\gamma$
are shown in FIG. \ref{fig1} and   FIG. \ref{fig2}, respectively.
When connect di-lepton line to one $\gamma$, operator $b\to s\gamma\gamma$ may
contribute to $B_s\to\ell^+\ell^-\gamma$.
Contributions from the such kind of diagrams with a  photon
attaching from internal charged lines to  $B_s\to\ell^+\ell^-\gamma$
are usually regraded as to be strongly suppressed by a factor $m_b^2/m_W^2$ thus can be neglected
safely~ \cite{Aliev97,Xiong01,LU06,Eilam97}. However, as pointed out in \cite{Dong}, 
the conclusion of this is correct,
but the explanation is not as what it is described.
Here we address the reason more clearly:  the
contribution from diagrams FIG. \ref{fig1} (a) and  FIG. \ref{fig1} (b) are not suppressed.
When applying effective vertex of $b\to s\gamma$ to describe $b\to s\gamma\gamma$ as
shown in FIG. \ref{fig2}, internal quarks in the effective vertex
are off-shell and such off-shell effects are also not suppressed.
We have proven  that the such two  non-suppressed effects in  FIG. \ref{fig1} and FIG. \ref{fig2}
cancel each other exactly \cite{Dong}.  Therefore we can use the effective operators
listed in Eq. (\ref{hami}) for on-shell quarks to calculate the total short distance contributions
 of $B_s\to\ell^+\ell^-\gamma$ in SM safely.

The Feynmann diagrams contributing to $B_s\to\ell^+\ell^-\gamma$ at parton
level can then be classified into three kinds as follows:
\begin{enumerate}
\item Attaching a real photon to any charged  external lines
in the Feynman diagrams of $b\to s\ell^+\ell^-$;
\item Attaching a virtual photon to any charged  external lines in
the Feynman diagrams of $b\to s\gamma$ with virtual photon into  lepton pairs;
\item Attaching two photon to any charged  external lines in the
Feynman diagrams of four-quark operators with one of two photon into
lepton pairs.
\end{enumerate}
Note that the third contribution is not considered in the previous studies except for Ref. \cite{Melikhov04} which 
is the focus of this paper and  will show the detail in the following. We also will  discuss these contributions seperatly.

\begin{figure}[htbp]
\begin{center}
\scalebox{0.8}{\epsfig{file=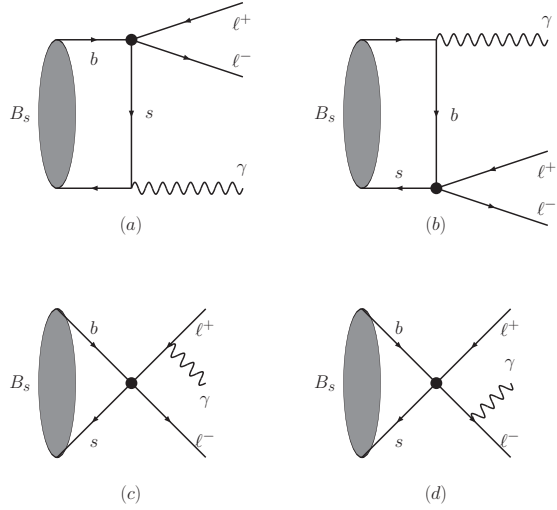}}
\caption{Feynman diagrams that contribute to the matrix elements
$B_s\to\ell^+\ell^-\gamma$ with the contribution  of $O_7,~O_9,~O_{10}$ in tree level.}
\label{fig3}
\end{center}
\end{figure}
\subsection{External real photon contributions}
The Feynman diagrams of the first kind of contributions are shown in FIG. \ref{fig3}.
As the contribution from the  FIG. \ref{fig3} (c) and (d) with photon
attached to external lepton lines, considering the fact that (i)
being a pseudoscalar meson, $B_s$ meson can only decay through axial current,
so the magnetic penguin operator $O_7$ 's contribution vanishes; (ii)
the contribution from operators $O_9,\ O_{10}$ has the helicity suppression
factor $m_{\ell}/m_{B_s}$, so for light lepton electron and muon,
we can neglect their contribution safely.
These diagrams in FIG. \ref{fig3} (a) and (b) are always regarded
as the dominant contributions, and they have
been considered by using the light cone sum rule~\cite{Aliev97,Xiong01}, the simple
constituent quark model~\cite{Eilam97},  and the B meson distribution
amplitude extracted from non-leptonic B decays~\cite{LU06}.
We rewrite the amplitude of $B_s\to \ell^+\ell^-\gamma$ at meson level as \cite{Wang:2012na}:
\begin{eqnarray}
A_{\rm I}&=&iN_cGee_d\frac{1}{p_{B_s}\cdot k}\biggl\{
\left[C_1 i \epsilon_{\alpha\beta\mu\nu}p^\alpha_{B_s}\varepsilon^\beta{k^\nu}
+C_2p_{B_s}^\nu(\varepsilon_\mu k_{\nu}- k_\mu\varepsilon_\nu)\right]
\bar{\ell}\gamma^\mu{\ell}\nonumber\\
&&+C_{10}\left[C_+i\epsilon_{\alpha\beta\mu\nu}p^\alpha_{B_s}\varepsilon^\beta{k^\nu}
+C_-p_{B_s}^\nu(\varepsilon_\mu k_{\nu}-k_\mu\varepsilon_\nu)\right]
\bar{\ell}\gamma^\mu\gamma_{5}\ell\biggl\}.
\label{ampp}
\end{eqnarray}
The form factors in Eq. (\ref{ampp}) are found to be:
\begin{eqnarray}
C_1 &=&C_+\left(C_9^{eff} -2\frac{m_{b}m_{B_{s}}}{q^{2}}C_{7}^{eff}\right), \nn\\
C_2 &=&C_{9}^{eff}C_{-}-2\frac{m_{b}m_{B_{s}}}{q^{2}}C_{7}^{eff}C_{+},
\label{formfa}
\end{eqnarray}
with  the constant $G=\alpha_{em} G_FV_{tb}V_{ts}^*/(\sqrt{2}\pi)$, and
\begin{eqnarray}
C_{\pm}=\int_{0}^{1}\left(\frac{1}{x}\pm\frac{1}{y}\right)\phi_{B_s}(x)dx.
\end{eqnarray} The expression in Eq. (\ref{ampp}) can be compared with Ref. \cite{LU06}.

\begin{figure}[htbp]
%\vspace{-0.5cm}
\begin{center}
\scalebox{0.8}{\epsfig{file=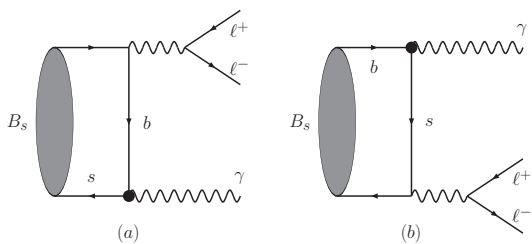}}
\caption{Feynman diagrams of $b\to s\gamma$ with virtual photon into
 lepton pairs .}
\label{fig4}
\end{center}
\end{figure}
\subsection{External virtual photon contributions}
The Feynman diagrams of the second kind of contributions are shown in FIG. \ref{fig4}.
Contributions from the  kind of diagrams  are always neglected~\cite{Aliev97,Xiong01}
or stated  to be negligibly small~\cite{LU06}. Note the $B_s$ meson wave functions used
in this work and Ref.~\cite{LU06}
are both from non-leptonic $B_s$ decays. However, as  mentioned in the introduction 
the authors of Ref. ~\cite{LU06}  did not present the expression of the contribution from
 FIG. \ref{fig4} and only stated the contribution is numerical  negligibly small.
But such  statement seems to be questionable, for that the pole of propagator of the
charged line attached by photon may enhance the decay rate greatly which make some
diagrams can not be neglected in the calculation.
In these two diagrams, photon of the magnetic-penguin operator is real,
thus its contribution to $B_s\to \ell^+\ell^-\gamma$ is different from the first kind
contributions. We get the amplitude \cite{Wang:2012na}:
\begin{eqnarray}
A_{\rm II}=i2N_cGee_dC_{7}^{eff}\frac{m_{b}m_{B_s}}{q^{2}}\frac{1}{p_{B_s}\cdot q}\overline{C}_+
\left[k_{\mu}q\cdot\epsilon-\epsilon_\mu k\cdot q
-i\epsilon_{\mu\nu\alpha\beta}\epsilon^{\nu} k^{\alpha}q^{\beta}\right]
\left[\bar{\ell}\gamma^{\mu}\ell\right],
\end{eqnarray}
with coefficients $\overline{C_+}$ obtained by a replacement:
\begin{eqnarray}
\overline{C}_+&=& C_+(x\to \bar{x}=x-z-i\epsilon;~y\to \bar{y}=y-z-i\epsilon)\nn\\
&=& N_B\int_{0}^{1}dx({\frac{1}{x-z-i\epsilon}}+{\frac{1}{1-x-z-i\epsilon}})
x^{2}(1-x)^{2}\exp\left[-\frac{m_{B_s}^2}{2\omega_{B_s}^2}x^{2}\right],
\label{ef}
\end{eqnarray}
where $z=\frac{q^{2}}{2p_{B_s}\cdot q}$ and the first, second term in  (\ref{ef}) 
denotes the contribution from FIG. \ref{fig4} (a) and (b) respectively. 
Note that the contribution from FIG. \ref{fig3} (a) is much larger than (b) 
since $m_{B_s}\ll\omega_{B_s}$ (see next section) which is very easily understood in sample
constituent quark model~\cite{Eilam97}, i.e., $\phi_{B_s}(x) =\delta (x-m_s/m_{B_s})$. 
However, the contributions from Fig.\ref{fig4} (a) and (b) are comparable, and  pole
in $\overline{C_+}$ corresponds to  the pole of the quark propagator when
it is connected by the off-shell photon propagator.
Thus the  $\overline{C_+}$ term may enhance
the decay rate of $B_s\to \ell^+\ell^-\gamma$  and its  analytic expression reads
\begin{eqnarray}
\overline{C}_+
&=&2N_{B_s}\pi i z^{2}(1-z)^{2}\exp\left[-\frac{m_{B_s}^2}{2\omega_{B_s}^2}z^{2}\right]\nn\\
&+&N_{B_s}\int_{0}^{1}dx({\frac{1}{x+z}}-{\frac{1}{1+x-z}})
x^{2}(1+x)^{2}\exp\left[-\frac{m_{B_s}^2}{2\omega_{B_s}^2}x^{2}\right]\nn\\
&-&N_{B_s}\int_{-1}^{1}\left(\frac{1}{\frac{1}{x}-z}+\frac{1}{1-\frac{1}{x}-z}\right)
\frac{dx}{x^{4}}(1-\frac{1}{x})^{2}
\exp\left[-\frac{m_{B_s}^2}{2\omega_{B_s}^2}\frac{1}{x^{2}}\right].
\end{eqnarray}

\begin{figure}[htbp]
\begin{center}
\scalebox{0.8}{\epsfig{file=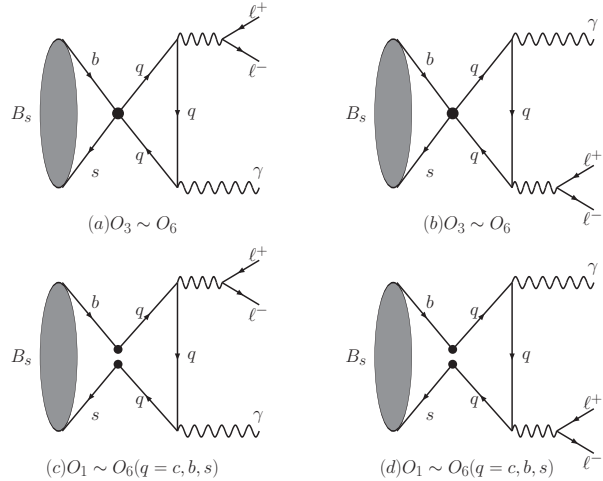}}
\caption{Feynman diagrams that contribute to the process  $B_s\to\ell^+\ell^-\gamma$ with
possible insertion of $O_1$ to  $O_6 $ in loop level .}
\label{fig5}
\end{center}
\end{figure}
\subsection{Quark weak annihilation  contributions}
Now we focus on the contributions from  the diagrams for the four-quark
operators which are not considered properly in previous works. 
The Feynman diagrams of the third kind  of contribution are shown in FIG. \ref{fig5}.
The operator $O_7$ is high order than four-quark operator $O_1 -O_6$,
thus the contribution of loop diagram with operator $O_1-O_6$ insertion
should be comparable with the tree level electro-weak penguin $O_7-O_{10}$
contributions listed above.  To calculate the leading order matrix
elements of $b\to s\gamma^\ast\gamma$ shown in FIG. \ref{fig5}, we express
the decay amplitude for  $b(p_b)\to s(p_s)\gamma^\ast(k_1)\gamma(k_2)$ as
\begin{eqnarray}
A_{\rm III}(b\to s\gamma^*\gamma )=i\frac{4G_F}{\sqrt{2}}V_{tb}V^*_{ts}\frac{e^2}{16\pi^2}\bar{s}(p_s)
 T_{\mu\nu}b(p_b)\epsilon^{\mu}(k_1) \epsilon^{\nu}(k_2),
\label{amptot}
\end{eqnarray}
where $p_{b,s}, k_{1,2}$  denotes the momentum of quarks and photon respectively,
$\epsilon$ is the vector polarization of photon. We split the tensor $T^{\mu\nu}$
into the momentum odd and even power terms  for simplification.
Keeping our physics goal in mind,  Without loss of generality we assume that
photon with momentum $k_1$ is virtual and drop the terms proportional to $k_2^\nu$
in the expressions. After straight calculation, we obtain
\begin{eqnarray}
T_{\mu\nu}^{\rm odd}(q)
&=&e_{q}^{2}\left\{\frac{1}{k_{1}k_{2}}\left[
i\epsilon_{\nu\alpha\beta\lambda}\gamma^{\lambda}\gamma_{5}k_{1}^{\alpha}k_{2}^{\beta}
\left[k_{1{\mu}}f_{2}(q)-k_{2\mu}f_{+}(q)\right]
+i\epsilon_{\mu\alpha\beta\lambda}\gamma^{\lambda}\gamma_{5}k_{1}^{\alpha}
k_{2}^{\beta}k_{1{\nu}} f_{+}(q)\right]\right.\nn\\
&&\left.+i\epsilon_{\mu\nu\alpha\beta}\gamma^{\beta}\gamma_{5}[k_{1}^{\alpha}f_{+}(q)
-k_{2}^{\alpha}f_{-}(q)] \right\},\\
T_{\mu\nu}^{\rm even}(q)&=&-2\frac{e_{q}^{2}}{m_q}\left\{
(k_{1\nu}k_{2\mu}-k_{1}k_{2}g_{\mu\nu})\left[f_{3}(q)+(1-\frac{4}{z_q})f'_{1}(q)\right]
+i\epsilon_{\mu\nu\alpha\beta}\gamma_{5}k_{1}^{\alpha}k_{2}^{\beta}f_{1}^{\prime}(q)\right\},
\label{Toddeven}
\end{eqnarray}
where $q$ denotes the quark in the internal line which two photons are attached and
and $e_q$ is the number of electrical charge of the quark.
The loop functions appearing in (\ref{Toddeven}) have forms as
 \begin{eqnarray}
f_{\pm}(q)&=&\frac{1}{2}+\frac{1}{z_q}\int_{0}^{1}\ln g(z_q,u_q,x)\frac{\mathrm{d}x}{x}
\mp\frac{u_q}{2z_q}\int_{0}^{1}\ln g(z_q,u_q,x),\nn\\
f_{1}(q)&=&\frac{1}{2}[f_+(q)+f_-(q)], \ f_2(q)=\frac{z_q}{2u_q}[f_+(q)-f_-(q)],\nn\\
f_{3}(q)&=&\frac{2}{z_q}-\frac{2u_q}{z_q^{2}}\int_0^1\mathrm{ln}g(z_q,u_q,x)\mathrm{d}x,\nn\\
g(z,u,x)&=&\frac{1-(u+z)x(1-x)}{1-ux(1-x)},
 \label{Loopfunc}
\end{eqnarray}
with  $f_1'(q)=\frac{1}{2}-f_1(q)$, $z_q=\frac{2k_{1}\cdot k_{2}}{m_q^{2}}$ and  $u_q=\frac{k_{1}^{2}}{m_q^{2}}$.
Writing the amplitude in  this ways, one can infer easily that for example,
when operators $O_{j}$ ($j=1,\dots,4$) are inserted, 
only $T_{\mu\nu}^{\rm odd}$ part can contribute to $b\to s\gamma\gamma$ 
while the process receives both parts contributions
as $O_{5,6}$ are inserted.  It is also easily obtained 
the similar result for the on-shell photons as in Ref.\ \cite{Cao01}
by setting $u_q=0$ .

With  the amplitude of $b\to s\gamma^*\gamma$ and $B_s$ wave function ready,
we  write the total contribution from FIG. 5 to exclusive decay of $B_s(p_{B_s})\to\gamma(k)\ell^+\ell^-$ as
\begin{eqnarray}
A_{\rm III}=-2ie\frac{f_{B_s}G}{q^2}
\sum\limits_{j=1}^6C_{j}(m_b)\left[T_{1}^{j}p_{B_s}^\nu(\epsilon_{\nu}k_{\mu}
-\epsilon_{\mu}k_{\nu})+T_{2}^{j}
i\epsilon_{\mu\nu\alpha\beta}p_{B_s}^{\alpha}k^{\beta}\epsilon^\nu\right]
[\bar{\ell}\gamma^\mu\ell],
\label{ampfig1}
\end{eqnarray}
with the form factors given by
\begin{eqnarray}
T_{1}^{1}&=&T_{1}^{2}= T_{1}^{3}=T_{1}^{4}=0;\nn\\
T_{2}^{1}&=&N_c T_{2}^{2}=N_{c}e_{u}^{2}f_{1}(\bar{z}_{c},t);\nn\\
T_{2}^{3}&=&N_{c}\left\{\sum\limits_{q=u,d,s,c,b}e_{q}^{2}f_{1}(\bar{z}_q,t)+e_{d}^{2}[f_{1}
(\bar{z}_b,t))+f_{1}(\bar{z}_s,t)]\right\}\nn\\
T_{2}^{4}&=&\sum\limits_{q=u,d,s,c,b}e_{q}^{2} f_{1}(\bar{z}_q,t)+e_{d}^{2}N_{c}\left[f_{1}(
\bar{z}_b,t))+f_{1}(\bar{z}_s,t)\right]\nn\\
T_{1}^{5}&=&=\frac{1}{N_c}T_{1}^{6}=2e_{d}^{2}\left\{\frac{1}{\bar{z}_{b^{1/2}}}
\left[f_{3}(\bar{z}_{b},t)
+(1-\frac{4\bar{z}_{b}}{1-t})(\frac{1}{2}-f_{1}(\bar{z}_{b},t))\right]-(b\to s)\right\};\nn\\
T_{2}^{5}&=&=-N_{c}\sum\limits_{q=u,d,s,c,b}e_{q}^{2}f_{1}(\bar{z}_q,t)
-2e_{d}^{2}\left[\frac{1}{\bar{z}_{b^{1/2}}}(\frac{1}{2}-f_{1}(\bar{z}_{b},t))+(b\to s)\right];\nn\\
T_{2}^{6}&=&-\sum\limits_{q=u,d,s,c,b}e_{q}^{2}f_{1}(\bar{z}_q,t)
-2N_ce_{d}^{2}\left[\frac{1}{\bar{z}_{b^{1/2}}}(\frac{1}{2}-f_{1}(\bar{z}_{b},t))+(b\to s)\right],
\label{Coefffig1}
\end{eqnarray}
where  $q^2$ in Eq.~(\ref{ampfig1}) is the invariant mass square of lepton pair.
The functions can be obtained directly  from (\ref{Loopfunc}) by redefined parameters
$\bar{z}_q=m_q^2/m^2_{B_s}$, $t=q^2/m^2_{B_s}$,
\begin{eqnarray}
f_1(\bar{z},t)&=&\frac{1}{2}+\frac{\bar{z}}{1-t}\int_{0}^{1}\frac{dx}{x}
\ln[\frac{\bar{z}-x(1-x)}
{\bar{z}-tx(1-x)}],\\
f_{3}(\bar{z},t)&=&\frac{2\bar{z}}{1-t}
\left\{1-\frac{t}{1-t}
\int_{0}^{1}{dx}\ln[\frac{\bar{z}-x(1-x)}
{\bar{z}-tx(1-x)}]\right\},
\label{Loopfunc1}
\end{eqnarray}
with explicit  formula needed in calculation given by
 \begin{eqnarray}
 \int_{0}^{1}\frac{dy}{y}\ln[1-v y(1-y)-i\epsilon]&=&\left\{
 \begin{array}{cc}
 -2\mathrm{arctan^{2}\sqrt{\frac{v}{4-v}}}& {\rm  for}~ v<4;\\
 -\frac{\pi^{2}}{2}-2i\pi\ln\frac{\sqrt{v}+\sqrt{v-4}}{2}+2\ln^{2}\frac{\sqrt{v}+\sqrt{v-4}}{2}& {\rm  for}~ v>4,\\
 \end{array}
 \right.,\nn\\
 \int_{0}^{1}dy\ln[1-v y(1-y)-i\epsilon]&=&
 -2+|1-x|^{1/2}\left\{
 \begin{array}{cc}
 \ln\left|\frac{1+\sqrt{1-x}}{1-\sqrt{1-x}}\right|-i\pi & {\rm  for}~ x=4/v<1;\\
 2\arctan\frac{1}{\sqrt{x-1}} &  {\rm  for}~ x=4/v>1.
 \end{array}\right.
 \label{Loopfunc2}
 \end{eqnarray}

From Eq. (\ref{ampfig1}) it is clear that the contribution from four-quark operators to
$B_s\to \gamma\ell^+\ell^-$ has the similar expression as that from
magnetic-penguin operator with real photon to $B_s\to \ell^+\ell^-\gamma$. Thus
the total matrix element for the decay $B_s\to\ell^+\ell^-\gamma$ including contributions
from three kinds of diagrams can be obtained easily by a shift to the form factors:
\begin{eqnarray}
\overline{C}_1&=&C_+\left[C_9^{eff}-\frac{2m_{b}m_{B_s}}{q^{2}}
C_7^{eff}\left(1+\frac{p_{B_s}\cdot k}{p_{B_s}\cdot q}\right)\right]
-2\frac{p_{B_s}\cdot k}{ q^2}\frac{f_{B_s}}{e_d}
\sum\limits_{j=1}^6C_j T_2^j,\\
\overline{C}_2&=&C_9^{eff}C_--\frac{2m_{b}m_{B_s}}{q^{2}}
C_7^{eff}C_+\left(1+\frac{p_{B_s}\cdot k}{ q^2}\right)
+2\frac{p_{B_s}\cdot k}{ q^2}\frac{f_{B_s}}{e_d}\sum\limits_{j=1}^6C_j T_1^j.
\label{formfactTOT}
\end{eqnarray}
Finally,  we get the
differential decay width versus  the photon energy $E_\gamma$,
\begin{eqnarray}
\frac{d\Gamma}{dE_\gamma}&=&\frac{\alpha_{em}^3{G^2_F}}{108\pi^4}|V_{tb}V^*_{ts}|^2
(m_{B_s}-2E_{\gamma})E_{\gamma}
%\nonumber\\&\times&
\left[|\overline{C}_1|^2+|\overline{C_2}|^2+C_{10}^2(|C_+|^2+|C_-|^2)\right].
\end{eqnarray}

\section{Results and discussions}\label{sec:num}
The decay branching ratios can be easily obtained by
integrating over photon energy. In the numerical calculations,
we use the following parameters~\cite{PDG2012}:
 $$\alpha_{em}=\frac{1}{137},~ G_F = 1.166\times 10^{-5} {\rm GeV}^{-2},
 ~m_b=4.2{\rm GeV},$$
 $$|V_{tb}|=0.999,~|V_{ts}|=0.04,~|V_{td}|=0.0084$$
 $$m_{B_s}=5.37{\rm GeV},~\omega_{B_s}=0.5,~f_{B_s}=0.24{\rm GeV},~\tau_{B_s}=1.47\times 10^{-12}s.$$
 $$m_{B_d^0}=5.28{\rm GeV},~\omega_{B_d} = 0.4,~f_{B_d} = 0.19{\rm GeV},~\tau_{B_d}= 1.53\times 10^{-12}s. $$
The ratios of $B_s\to \gamma\ell^+\ell^-$ are shown in Table \ref{table} together with
 results of $B_{d,s}\to\gamma\ell^+\ell^-$  from this work and our previous research for comparison.
The errors shown in the Table~\ref{table} comes from the heavy meson wave function, by
varying the parameter $\omega_{B_d}=0.4\pm 0.1$, and
$\omega_{B_s}=0.5\pm 0.1$ \cite{LU06}.
Note that, the predicted branching ratios receive  errors from
many parameters, such as parameters meson decay constant, meson and quark masses.
\begin{table}[htbp]
\caption{Comparison of branching ratios in unit of $10^{-9}$  with previous calculations}
\begin{center}
\begin{tabular}{c|c|c|c}
\hline\hline
Branching Ratios &\multicolumn{3}{c}{Results}\\
\hline
($\times 10^{-9}$) &The first kind of diagrams& The first two kind of diagrams & Included all diagrams\\
\hline
$B_s\to \gamma\ell^+\ell^-$ & $5.16_{-1.38}^{+2.42}$  & $10.2_{-2.51}^{+4.11}$  & $17.36_{-2.63}^{+4.55}$ \\
$B_d^0\to \gamma\ell^+\ell^-$ & $0.21_{-0.06}^{+0.14}$ & $0.40_{-0.13}^{+0.26}$ & $0.53_{-0.12}^{+0.26}$ \\
\hline\hline
\end{tabular}
\label{table}
\end{center}
\end{table}
%=================================================

From the numerical results we conclude that
unlike in decay $B\to X_s\gamma\gamma$ where the four-quark operators just contribute a
few percent to the branching ratio \cite{Cao01}, our numerical result shows that contribution
from  the four-quark operators to $B_s\to \gamma\ell^+\ell^-$ is large.
It can be understood as follows:
\begin{enumerate}
\item As pointed out in Ref.~\cite{LU06},  the radiative
      leptonic decays are  very sensitive probes in extracting the heavy
      meson wave functions;
\item Values of the Wilson coefficients $C_j(m_b)$ ($j=3,\dots,6$) are at order of $10^{-2}$,
      indicating that contribution to the corresponding operators
      via $T^j$ is less important compared with those from $O_{1,2}$;
\item From Eq. (\ref{Coefffig1}), one can infer easily that the four-quarks 
      contribution  to the form factors in (\ref{formfactTOT}) have coefficient
      $(N_cC_1+C_2)T_2^2f_{B_s}/e_d$ while the contribution from
     magnetic-penguin operator with real photon,  $m_{b}m_{B_s}/(p_{B_s}\cdot q)C_7^{eff}C_+$.
     Note $(N_cC_1+C_2)/e_d$ and $C_7^{eff}$ can be comparable  and have the same sign
      in $\overline{C}_1$ and opposite sign in $\overline{C}_2$. However, with $T_1^j=0$ for $j=1,2$
     thus comparable contribution studied in this work and in Ref. \cite{Wang:2012na} is expected,
     leading to enhancement of branching ratios of $B_s\to\gamma\ell^+\ell^-$ when new
      diagrams are taken into account.
\item  The predicted short-distance contributions from quark weak annihilation as well as
     the magnetic-penguin operator  with real photon to
     the exclusive decay are large, and  the branching ratios of $B_s\to \ell^+\ell^-\gamma$
     are  enhanced nearly by a factor 3 compared with that only contribution
     from magnetic-penguin operator with virtual photon
     and up to $1.7\times 10^{-8}$, implying the search
     of $B_s\to \ell^+\ell^-\gamma$  can be achieved in near future.
\item Due to the large contributions from magnetic-penguin operator  
      with real photon and quark weak annihilation , the form factors for matrix elements 
      $\langle\gamma|\bar{s}\gamma_{\mu}(1-\gamma_5) b|B_s\rangle$ and 
      $\langle\gamma|\bar{s}\sigma_{\mu\nu}(1\pm\gamma_5)q^\nu b|B_s\rangle$ as a 
     function of dilepton mass squared $q^2$ are  complex and not  as simple as
     $1/(q^2-q_0^2)^2$ where  $q_0^2$ is constant \cite{Eilam95}. 
     The $B_s\to \gamma$  transition form factors predicted in this works have also some differences
     from those in Ref. \cite{Melikhov04,Kruger03,Nikitin11}. For instance,
     Ref. \cite{Kruger03} predicted the form factors $F_{TV}(q^2,0)$, $F_{TA}(q^2,0)$
     induced by tensor and pseudotensor currents
     with direct emission of the virtual photon from quarks
     are only equal at maximum photon energy, whereas the corresponding 
      formula in this work have the same expression  
     as $-\frac{e_dN_cm_{B_s}}{p_{B_s}\cdot k} C_+\propto 1/(q^2-q_0^2)$ in 
     Eq. (\ref{ampp}). Furthermore, the form factors  are larger than previous predictions.
\end{enumerate}

To clarify things more clear, we think it is necessary to  present a few more comments about the
calculation of Ref. \cite{Melikhov04}, as mentioned in introduction.
In order to estimate the contribution of direct
emission of the real photon from quarks, the authors of Ref. \cite{Melikhov04} calculated the form factors
$F_{TA,TV} (0, q^2)$ by including the short-distance contribution in $q^2\to 0$ limit
and additional long-distance contribution from the resonances of vector mesons
such as  $\rho^0$, $\omega^0$ for $B_d$ decay and $\phi$  for $B_s$ decay.
Obviously, this means the short-distance contributions
were  not appropriately taken into account. Moreover,
if $F_{TA,TV} (0, q^2)=F_{TA,TV} (0, 0)$ stands for the short distance
contribution, it seems to double counting since in this case
photons emitted from magnetic-penguin vertex and quark lines directly
are not able to  be distinguished.

We also note that for contribution from the weak annihilation the authors of Ref. \cite{Melikhov04}
only took into account $u$ and $c$ quarks in the loop by axial anomaly as the long distance contribution,
they  concluded that the anomalous contribution is
suppressed by a power of a heavy quark mass. We believe that only anomalous contribution
to account contribution from the weak annihilation is insufficient. 
Our numerical result shows that the contributions from weak-annihilation diagrams
are large and can not be neglected.

\section{Conclusion}\label{sec:con}
In summary, we evaluated  short distance calculation of the rare decays
$B_s\to \gamma\ell^+\ell^-$ in the SM,  including contributions from
all four kinds of diagrams. We focus  on the contribution from
four-quark operators which are not taken into account properly in previous researches.
We found that the contributions are large, leading to  the branching ratio of
 $B_s\to \ell^+\ell^-\gamma$ being  nearly enhanced by a factor 3.
In the current early phase of the LHC era, the exclusive modes with muon final states are
among the most promising decays. Although there are some theoretical challenges in calculation
of the hadronic form factors and non-factorable corrections, with the predicted branching
ratio at order of  $10^{-8}$,   $B_s\to\mu^+\mu^-\gamma$ can be expected as
 the next goal after  $B_s\to \mu^+\mu^-$ since the final states can be
identified easily and the branching ratios are large.  Experimentally, 
$B_s\to \mu^+ \mu^-\gamma$ mode is
one of the main backgrounds to $B_s \to\mu^+\mu^-$, and thus
it is already taken into account in $B_s \to\mu^+\mu^-$ searches \cite{Aaij:2012nna}.
 Our predictions for such processes can be tested in the LHC-b and B factories in near future.
\begin{acknowledgments}
This work was supported in part by the NSFC No. 11005006, 11172008.
\end{acknowledgments}

\end{document}